\documentclass[aps,pre,onecolumn,showpacs,floatfix]{revtex4}
\usepackage{amssymb,amsmath,epsfig,graphicx}

\usepackage{caption}
\usepackage{dcolumn}

\usepackage{amssymb,graphicx}

\begin{document}

\title{Optimal operation of feedback flashing ratchets}

\author{M Feito$^1$, F J Cao$^{1,2}$}
\affiliation{$^1$ Departamento de F\'{\i}sica At\'omica, Molecular y
Nuclear, Universidad Complutense de Madrid, 
Avenida Complutense s/n, 28040 Madrid, Spain.}
\affiliation{$^2$
LERMA,
Observatoire de Paris, Laboratoire Associ\'e au CNRS UMR 811 2,
61, Avenue de l'Observatoire, 75014 Paris, France.
}
\email{feito@fis.ucm.es}
\email{francao@fis.ucm.es}

\begin{abstract}
Feedback flashing ratchets are thermal rectifiers that use
information on the state of the system to operate the switching on and
off a periodic potential. We discuss different strategies for this operation
with the aim of maximizing the net flux of particles in the collective version
of a flashing ratchet consisting on $N$ overdamped Brownian particles. We show
the optimal protocols for the one-particle ratchet and for the  collective
ratchet with an infinite number of particles. Finally we comment on the
unsolved problem of the optimal strategy for any other number of particles.
\end{abstract}



\maketitle


\section{Introduction}
\begin{figure}
\begin{center}
\includegraphics [scale=0.7] {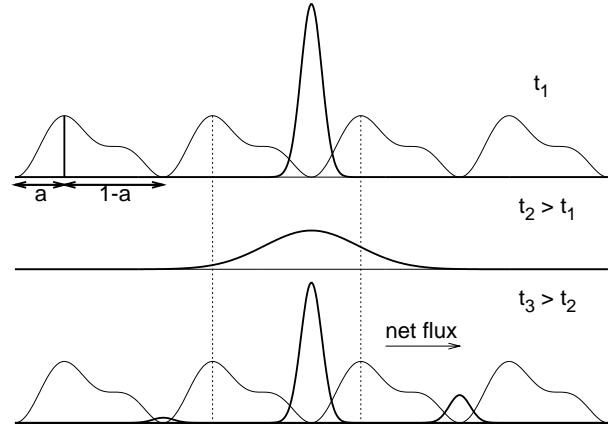}
\end{center}
\caption{Illustration of a flashing ratchet operation. The ratchet potential
  depicted has spatial period $1$ and asymmetry $a=1/3$. The thick lines
  represent the probability density of a Brownian particle initially at one
  of the minima of the potential (top). When the potential is
  switched off the particle begins to diffuse isotropically
  (middle). After the potential is switched on again a net flux to
  the right is obtained due to the asymmetry of the potential 
  (bottom).
}\label{figure_1}
\end{figure}

Brownian motors or ratchets are rectifiers of thermal fluctuations that 
induce direct transport without an a priori bias~\cite{rei02}. The breaking of
thermal equilibrium and of certain time-space symmetries are necessary
conditions to manage direct transport~\cite{den08,fla00,yev01}.
Ratchets are relevant from the theoretical point of
view to study non-equilibrium processes, and from a practical point of
view due to their applications in nanotechnology and
biology~\cite{rei02,lin02,kay07}. A prototypical example of a ratchet system
is the so-called flashing ratchet, which is based on switching on and off a
periodic potential~\cite{ajd93,ast94}. The spatial asymmetry of the
potential ensures a direct transport, but also a symmetric potential under an
intrinsically asymmetric operation for the switching can achieve this task.
Figure~\ref{figure_1} illustrates the operation of a flashing ratchet.
\par

Control strategies such as those obtained with a time periodic or random
switching are open-loop control policies, as they do not use any information
about the particle distribution of the system to operate. 
Some studies deal with the optimal modulation in time of ratchet potential
to maximize the performance of these open-loop systems~\cite{tar98,sch07}. 
These works uses a variational approach to optimize the velocity induced by
deterministic modulation in time of the potential.

On the other hand, it has been shown that a significant increase for the net
flux in a flashing ratchet can be obtained if feedback on the state of the
system is used by the protocol that switches on and off the ratchet
potential~\cite{cao04} (i.e., taking into account the positions of
the Brownian particles before the switching operation). This can improve the
technological applications of ratchets. Experimental
implementations of these feedback (= closed-loop control) flashing ratchets
have been proposed~\cite{fei07,cra08a,cra08}, and the 
realization of such devices is currently under way~\cite{cra08,lop08}. In
addition, feedback ratchets have been suggested as a mechanism to explain the
stepping motion of the two-headed kinesin~\cite{bie07} and are also present in
other chemically-driven molecular motors~\cite{zho96,ast98,alv08}.
\par

In this context, determining the optimal protocol for the operation of 
collective feedback ratchets is a relevant question that has
only received partial answers. In the present work we review and compare some
control strategies designed with the aim of maximizing the net flux of
particles. We show the optimal protocol for the one-particle
flashing ratchet and for the 
limit case of an infinite number of particles. In the following section we
present the Langevin equations standing for the feedback ratchet. After that,
we revise the instant maximization protocol (Sec.~\ref{sec:p1}), the threshold
protocol (Sec.~\ref{sec:p2}), and the maximal net displacement protocol
(Sec.~\ref{sec:p3}).  We finally discuss the main unsolved questions related
with the optimal operation of these systems (Sec.~\ref{sec:upon}).

\section{Collective flashing ratchet}
A collective flashing ratchet can be modeled by $N$ Brownian particles at
positions $\{x_j(t)\}$ that satisfy the overdamped Langevin equations
\begin{equation}\label{collective}
\gamma \dot x_i(t)=\alpha(\{x_j(t)\},t)F(x_i(t))+\xi_i(t);\qquad i=1,\dots,N.
\end{equation}
Here, $F(x)=-V^\prime(x)$, with $V(x)$ the so-called ratchet
potential, which is usually spatially periodic, $V(x)=V(x+L)$, and has broken
symmetry $x\to -x$. In these equations $\xi_i(t)$ stand for 
Gaussian white noises of zero mean and variance $\langle
\xi_i(t)\xi_j(t^\prime)\rangle =2\gamma k_B T\delta_{ij}\delta(t-t^\prime)$,
with $\gamma$ the viscous friction coefficient.
The function $\alpha(t)$ implements the action
of a controller that switches on the ratchet potential ($\alpha=1$) or switches
it off ($\alpha=0$). Thanks to this switching, detailed balance symmetry is
broken and the system can act as a Brownian motor. Note that, even for
symmetric potentials $V(x)=V(-x)$, there can be ratchet effect provided the
control policy induces the required asymmetry.

A common ratchet potential is
\begin{equation}\label{typicalpot}
V(x)=V(x+L)=\frac{2V_0}{3\sqrt{3}}\left[ \sin\left(\frac{2\pi
      x}{L}\right)+\frac{1}{2} \sin\left(\frac{4\pi x}{L}\right)\right],
\end{equation}
which has height $V_0$ and asymmetry parameter $a=1/3$ (with $aL$ the distance
between a minimum and the next maximum). This potential has
been used in figure~\ref{figure_1} to illustrate the flashing ratchet effect.
\par

It is worthwhile to point out that there is no dependence of the average flux
with the number of particles for open-loop control policies, as the Langevin 
equations are decoupled. However, for closed-loop ratchets the feedback
strategy introduces a coupling between the particles dynamics, and the
center-of-mass velocity does depend on $N$.

\section{The instant maximization protocol}\label{sec:p1}

In the \emph{instant maximization of the velocity protocol}~\cite{cao04} the
control policy depends on the sign of the net force per particle at each
instant of time. More specifically, the potential is switched on if the net
force the particles would feel with the potential `on':
\begin{equation}\label{netforce}
f(t)=\frac{1}{N}\sum_{i=1}^N F(x_i(t)),
\end{equation}
is positive, and the potential is switched off otherwise. Thus
\begin{equation}
  \alpha(t)=\Theta(f(t)), 
\end{equation}
with $\Theta$ the Heaviside function [$\Theta (x)=1$ if $x>0$, else $\Theta
(x)=0$]. 
\par

For a one-particle ratchet this protocol gives the maximum possible flux,
i.e., it is the optimal protocol for $N=1$. Let
us call $x$ the cyclic coordinate of the particle (modulo $L$). Then, the
instant maximization protocol just consists of
switching on the potential whenever the particle is in a region with a positive
slope of the potential, $x\in (aL,L)$, or switching off the potential
whenever the particle is in a region with a negative slope, $x\in
(0,aL)$. Hence 
the system can be reinterpreted as a Brownian particle that freely diffuses in
the flat regions $(aL,L)$ and `slides down' in the biased
regions $(0,aL)$. Now it is clear that the instant maximization protocol is the
optimal strategy for the one-particle ratchet, as any changing in the
prescription $\alpha(x(t))$ would imply either changing the flat regions to
uphill potential barriers or change the downhill regions to plateaus. 
A similar argument allows us to proof the more general statement that the 
instant maximization protocol beats any general protocol $\alpha(x(t),t)$, 
even with an explicit dependence in $t$ and $\alpha$ varying in the range 
$[0,1]$ (pulsating ratchets). From the Langevin equation (\ref{collective}), 
with $N=1$, the average steady state velocity is formally computed as 
\begin{equation}\label{dem1}
  \langle \dot x \rangle=\lim_{\tau\to\infty}
  \frac{1}{\gamma\tau}
  \int_0^\tau
  \alpha(x(t),t)F(x(t))\;dt.
\end{equation}
On the other hand, the effective force including the action of the controller
in the instant maximization operation is $\Theta(F(x(t)))F(x(t))$, so that at
each instant of time $t$ the inequality 
\begin{equation}\label{dem2}
  \alpha(x(t),t)F(x(t))\leq \Theta(F(x(t)))F(x(t)),
\end{equation}
holds for any control $0\leq \alpha(x(t),t)\leq
1$, as the maximum value for a positive (negative) force is obtained when the
control parameter is $\alpha=1$ ($\alpha=0$). We finally combine
inequality~(\ref{dem2}) and equation~(\ref{dem1}), which proves the optimality
of the instant maximization protocol.

\par

However, for a large number of particles the system dynamics gets
trapped with the potential `on' or `off', as the fluctuations of the magnitude
of $f(t)$ ---which trigger the switches--- become
smaller~\cite{cao04}. Therefore useless waiting times with the particle
distribution near the equilibrium are 
wasted after the next switching. For a thousand particles the instant
maximization protocol gives a lower flux than a simple periodic
switching strategy that does not receives any feedback from the
system. In the end, the flux goes to zero as $N$ increases; see
figure~\ref{fig:figure_2}.
\par

\section{The threshold protocol}\label{sec:p2}
The undesired trapping of the many-particle dynamics is settled in the
\emph{threshold protocol}~\cite{din05,fei06}, in which the potential
switchings are forced, provided the magnitude of the net force is below certain
threshold values and the system is in an unproductive state near
equilibrium.
\par

\begin{figure}
\begin{center}
\includegraphics[scale=0.7]{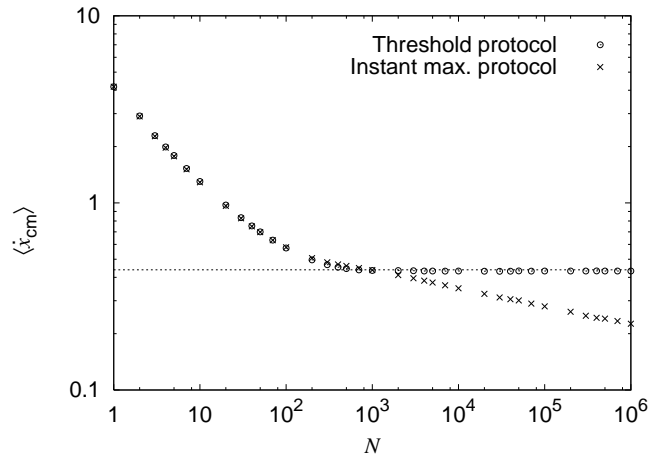}
\end{center}
\caption{\label{fig:figure_2}
Stationary center-of-mass velocity as a function of the number of
particles for the instant maximization protocol and for the threshold
protocol  with optimal thresholds for $N\to\infty$. The dashed line stands for
the periodic switching protocol with optimal periods. Units: $L=1$,
$\gamma=1$, $k_BT=1$. We have used the ratchet potential~(\ref{typicalpot})
with potential height $V_0=5$. For this system the optimal thresholds for
$N\to\infty$ are $u_{\mbox{\footnotesize{on}}}\simeq 0.6$ and
$u_{\mbox{\footnotesize{off}}}\simeq -0.4$. The estimated optimal semiperiods
for the on and off times of the periodic protocol are $0.06$ and $0.05$,
respectively.}
\end{figure}

Let us call $u_{\mbox{\footnotesize{on}}}\geq0$ and $u_{\mbox{\footnotesize{off}}}\leq 0$ the thresholds
values that induce switchings off and on when the decaying long tails of
$f(t)$ are still positive or negative, respectively. Then,
the \emph{threshold control} is given by
\begin{equation}\label{alfa}
\alpha(t)=
\begin{cases}
1& \mbox{if } f(t)\geq u_{\text{on}},\\
1& \mbox{if } u_{\text{off}}<f(t)<u_{\text{on}} \mbox{ and } \dot f_{\text{exp}}(t)\geq 0,\\
0& \mbox{if } u_{\text{off}}<f(t)<u_{\text{on}} \mbox{ and } \dot f_{\text{exp}}(t)< 0,\\
0& \mbox{if } f(t)\leq u_{\text{off}}.
\end{cases}
\end{equation}
The condition over the sign of the derivative of the net force ensures that
the switchings are only enforced when the system is really relaxing to
equilibrium. This derivative is obtained by applying the It\^o chain rule to
the stochastic magnitude $f(t)$~\cite{din05,fei06}:
\begin{equation}
\dot f_{\mbox{\footnotesize{{exp}}}}(t) =\frac{1}{\gamma N}\sum_i
\alpha(t)F(x_i(t))F^\prime(x_i(t)) +\frac{k_B T}{\gamma N}\sum_i
F^{\prime\prime}(x_i(t)).
\end{equation}
The instant maximization protocol ---optimal for $N=1$--- is recovered by
taking zero threshold $u_{\mbox{\footnotesize{{on}}}}= u_{\mbox{\footnotesize{{off}}}}=0$. On the other
hand, for an adequate value of the thresholds, this new
strategy gives the same flux as the optimal periodic, open-loop
switching protocol for an infinite number of particles. Note that for
$N\to\infty$ no appreciable advantage over the optimal periodic 
switching can be obtained by using a feedback scheme, as only one 
degree of freedom is performed in the control (switching on and off).
Therefore, the family of threshold protocols succeeds in getting the optimal
control both for the one particle case and for the infinite particle case,
yielding in between fluxes greater than or equal to open-loop protocols for all
numbers of particles. In fact, the specific thresholds
$u_{\mbox{\footnotesize{{on}}}}$ and $u_{\mbox{\footnotesize{{off}}}}$ that are
optimal for $N\to\infty$ also give a high value for the flux close to the
value corresponding to the best thresholds for finite
$N$~\cite{fei06}. Figure~\ref{fig:figure_2} shows the flux dependence with the
number of particles for the instant maximization protocol and the threshold
protocol with optimal thresholds for $N\to\infty$. The flux is also compared
with the $N$-independent value for the best open-loop, periodic switching
protocol.

\section{The maximal net displacement protocol}\label{sec:p3}
The previous feedback control strategies are based on the distribution of the
forces $\{F(x_i)\}$. But other choices are also possible.  For instance, in
the so-called \emph{maximal net displacement protocol}~\cite{cra08} the
control parameter depends on the particle distribution as follows,
\begin{equation}
  \alpha(t)=\Theta(d(t)); \quad d(t)=\sum_{i=1}^N (x_i(t)-x_0),
\end{equation}
where $x_0$ is the mean of a Gaussian distribution at equilibrium in the
ratchet potential $V(x)$.
This new protocol was numerically found~\cite{cra08} to slightly beat the
instant maximization protocol for collective ratchets of two and three
particles and potential heights $\gtrsim 30 k_BT$. However, it performs worse
for other numbers of particles or smaller potential heights.

\section{Discussion and open questions}\label{sec:upon}

The previous results provide some partial answers to the question of the
optimal operation of feedback ratchets to maximize the flux.
The appropriate introduction of feedback policies in collective flashing
ratchets improves the performance of the system for a finite number of
particles. In particular, the instant maximization protocol~\cite{cao04} is the
optimal operation 
for the one-particle ratchet, as we have proven. However, this `greedy'
strategy is not optimal 
for the collective version of the ratchet. This kind of short-range
maximization has also been reported not to be optimal in the so-called
paradoxical games~\cite{din03,cle04}. On the other hand, in the limit of an 
infinite number of particles no advantage over open-loop strategies can be 
achieved by using any feedback in the system. Our results also show that the 
threshold protocol~\cite{din05,fei06} with proper threshold values 
is optimal for $N=1$ (zero thresholds) and $N\to\infty$, and it can give large
values of the flux for intermediate number of particles. It is interesting to
note that an exact optimization study in 
the framework of a discrete ratchet-like systems revealed that the optimal 
protocol is indeed a kind of threshold operation~\cite{cle04}.
\par

Despite the advance that these partial answers represent, the fundamental
problem ---which is the optimal protocol for a collective feedback flashing
ratchet--- is still unsolved. A systematic study about the maximization of the
flux in these systems using either the stochastic version of the Pontryagin
maximum principle or the stochastic version of the Bellman's dynamic
programming~\cite{yon99} is still lacking.  On the other hand, we point out
that adding another external perturbation to the feedback flashing ratchet can
enlarge the flux of the system, as it happens, for instance, when a feedback
flashing ratchet is rocked with proper amplitude and frequency~\cite{xxx}.
We also point out that, to the best of our knowledge, there is still no study
of the effects of inertia in feedback ratchets. These studies will also broaden
the implications and applications of feedback ratchets.
\par

Finding the optimal protocol of feedback ratchets would be an important
theoretical task, as these systems are nothing other than Maxwell's
demons~\cite{lef03}; thus it would give a paradigmatic example to illustrate 
the maximum performance that can be attained by using a certain amount of
information obeying thermodynamic restrictions.  On the other hand, there is
also an increasing interest in feedback flashing ratchets as nanotechnological
devices (see~\cite{cra08} for a description of two set-ups capable of
implementing a collective feedback flashing ratchet; one of these experiments
has been very recently done~\cite{lop08}). In addition, the
maximization of the instant velocity, which is optimal for $N=1$ as we have
proved, has allow one to get an insight into the motion of a linear, two
headed, processive molecular motor~\cite{bie07}. A deep
knowledge of the maximum flux that can be attained in feedback flashing
ratchets could also help to get insight into other biological systems, such as
molecular motors operating in a collective way~\cite{jul97}.

\section*{Acknowledgments}
We acknowledge financial support from MCYT (Spain) through the Research
Project FIS2006-05895, from the ESF Programme STOCHDYN, and from 
UCM and CM (Spain) through CCG07-UCM/ESP-2925.

\end{document}